# Where in the world are you?

## Geolocation and language identification in Twitter

Mark Graham, Scott A. Hale, Devin Gaffney

Oxford Internet Institute, University of Oxford

All authors contributed equally to this paper.

Abstract

The movements of ideas and content between locations and languages are unquestionably crucial concerns to researchers of the information age, and Twitter has emerged as a central, global platform on which hundreds of millions of people share knowledge and information. A variety of research has attempted to harvest locational and linguistic metadata from tweets in order to understand important questions related to the 300 million tweets that flow through the platform each day. However, much of this work is carried out with only limited understandings of how best to work with the spatial and linguistic contexts in which the information was produced. Furthermore, standard, well-accepted practices have yet to emerge. As such, this paper studies the reliability of key methods used to determine language and location of content in Twitter. It compares three automated language identification packages to Twitter's user interface language setting and to a human coding of languages in order to identify common sources of disagreement. The paper also demonstrates that in many cases user-entered profile locations differ from the physical locations users are actually tweeting from. As such, these open-ended, user-generated, profile locations cannot be used as useful proxies for the physical locations from which information is published to Twitter.

Microblogging services such as Twitter allow researchers, marketers, activists and governments unprecedented access to digital trails of data as users share information and communicate online. Patterns of information exchange on platforms that rely on user-generated content have been used recently in scholarly research about community (Gruzd, Wellman, and Takhteyev 2011), information diffusion (Romero, Meeder, and Kleinberg 2011), politics (Bruns and Burgess 2011), religion (Shelton, Zook, and Graham 2013), crisis response (Zook et al. 2010; Palen et. al. 2011), and many other topics. Such data are also important to governments and marketers seeking to understand trends and patterns ranging from customer/citizen feedback to the mapping of health pandemics (Graham and Zook 2011). Twitter in particular with its large and international user base (there are now over 350 million users on the platform) has been the source of much scholarly research.

Content passed through Twitter remains decontextualized, however, unless we find ways to reattach it to geography. In other words, we don't just want to know what is said, but we also want to know where it is said and to whom it is said. As such, the attributes of language and location are crucial for understanding the geographies of online flows of information and the ways that they might reveal underlying economic, social, political and environmental trends and patterns. Yet, both language and location are challenging to deduce in the short messages that pass through Twitter, and no well accepted methodology for their extraction and analysis has been articulated. This point is especially salient because of the increasing number of studies, journalistic accounts, and real-world applications that rely on harvested locational and language data from Twitter. Therefore, in order to provide a useful starting point for future reach on Twitter (and indeed other micro-blogging platforms), this paper compares several approaches to working with geographic information in Twitter in order to better understand the strengths and limitations of each.

The short size of posts (140 characters on Twitter) presents a challenge to accurate language identification due to the fact that most language identification algorithms are trained on larger sized documents (Carter, Tsagkias, and Weerkamp 2011). In addition, the style of writing on Twitter using abbreviations and acronyms complicates language classification. In many instances, researchers have simply relied on the user interface language of a user's account or used an off-the-self language detection package without consideration of its suitability for use on short, informal text phrases. The disagreement of several studies on the most used languages in Twitter (Honeycutt and Herring 2009; Semiocast 2010; Hong, Convertino, and Chi 2011; Takhteyev, Gruzd, and Wellman 2011) highlights the difficulty of language detection . All four studies agree English is the most used language, but give percentages ranging from 50 percent (Semiocast 2010) to 72.5 percent (Takhteyev, Gruzd, and Wellman 2011). The purpose of our work is not to study the prominence of different languages on the platform, but is rather to highlight important methodological issues related to language identification in order for future research to more critically engage in geolinguistic analyses.

Accurately determining location in messages sent through Twitter is also a significant challenge. The most apparent method is to consider the profile information that is directly provided by a user (e.g. the text "Washington, DC" in Figure 1) in response to an account set-up question: "Where in the world are you?" However, this question, which allows users to input any text string to describe their location (referred to in this paper as 'profile location'), is often hard to geolocate correctly (the open-ended text could just as easily say "Edinburgh, Scotland," "Barad-dûr, Mordor, Middle-earth," or simply "here"). High error rates, missing data and non-standardized text in profile locations have forced some researchers wishing to employ this geographic data to use smaller samples and labor-intensive manual coding of profile locations (e.g. Takhteyev, Gruzd, and Wellman 2011).

**Figure 1**: Screenshot from Barack Obama's Twitter profile page.

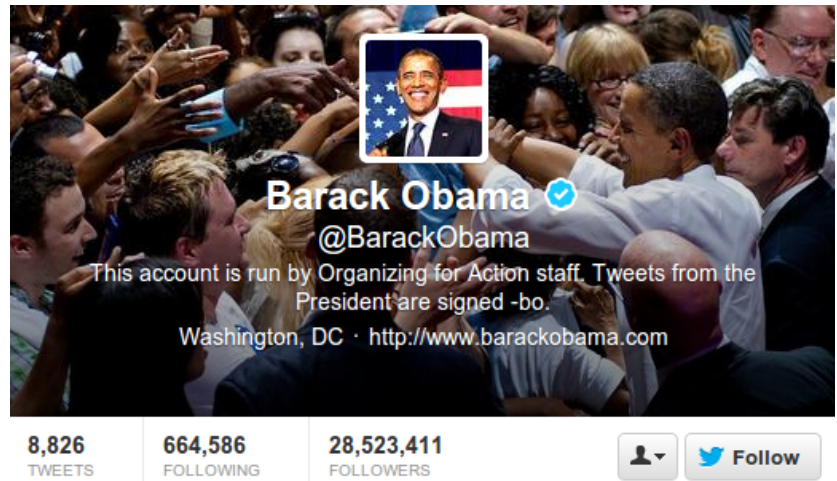

An alternate approach that some researchers have adopted is to narrow their samples to only use geocoded tweets. Depending on user's privacy settings and the geolocation method used, these tweets have either an exact location specified as a pair of latitude and longitude coordinates or an approximate location specified as a rectangular bounding box. This type of geographic information (referred to in this paper as 'device location') represents the location of the machine or device that a user used to send a message on Twitter. More precisely, the data are derived from either the user's device itself (using the Global Positioning System [GPS]) or by detecting the location of the user's Internet Protocol (IP) address. Precise coordinates are almost certainly from devices with built-in GPS receivers (e.g. phones and tablets). Bounding boxes, however, can result from privacy settings applied to GPS data or from GeoIP data.

Irrespective of these limitations, device locations are challenging for users to manually manipulate, and, because they are structured data, are easily interpreted by computers. However, only a small portion of users publish geocoded tweets, and it is unlikely that they form a representative sample of the broader universe of content (i.e. the division between geocoding and non-geocoding users is almost certainly biased by factors such as social-economic status, location, education, etc.). From a sample of 19.6 million tweets collected by the authors (these data were collected using Twitter's 'statuses/sample stream' collection method with 'spritzer access') over nineteen days in June 2011, only 0.7 percent of tweets contained structured geolocation information. As such, the extremely low proportion of information with attached device locations means that researchers either have to work with data that are likely highly skewed or devise effective methods to work with the profile location that is attached to all of the tweets that do not contain explicitly geocoded device location information.

This paper deals with these gaps of knowledge related to language and location in two primary ways. First, it explores the accuracy of a range of language detection methods on tweets: which, by, definition, are short and often contain informal phrasings and abbreviations. It identifies common sources of errors and compares performance over four research locations, each comprising a large variety of languages. Second, it compares various location information within tweets (profile location, device location, timezone information) and the accuracy with which geolocation algorithms can interpret the free-form profile location information.

In performing this work, we are able to refine methods that can be employed to map and measure the geolinguistic contours of people's information trails on Twitter. Doing so will ultimately allow future work to build on this research in order to create more accurate and nuanced understandings of the clouds of digital information that overlay our planet.

**Related Work**

A variety of methods have been employed in looking at Twitter's geolinguistic contours. Hong et al. (2011) used two automated tools to determine the language of a tweet: LingPipe and the Google Language Application Programming Interface (API), while Semiocast (2010) use an internal proprietary tool. Carter et al. (2011) and Gottron and Lipka (2010) discuss several of the challenges with language identification on short texts, the largest being that most language detection algorithms have been developed and trained on full documents that are longer and better formulated than the short text snippets that pass through Twitter. Carter et al. (2011) focus on microblog posts and develop two approaches (priors) to enhance performance: a link-based approach to consider the language of linked-to content and a blogger-based approach to aggregate tweets on a per account basis to form a larger document to classify. They find both approaches improve accuracy, but still leave room for further improvement. Hale (2012a) used the Compact Language Detection (CLD) kit, part of Google Chrome, for detecting the language of blogs in conjunction with the presence of certain keywords. He found these two methods in combination to be 95 percent accurate on a sample of 965 blogs about the Haitian earthquake. The CLD has since been used for in creating visualizations of language on Twitter (Fischer 2011), but its accuracy has not yet been evaluated for short posts passed through Twitter.

While geographic metadata in device locations (i.e. precise coordinates) are unlikely to be subject to much debate about their validity, the self-reported profile location field in a user's profile is problematic because of its unstructured nature. However, it remains that the usage of profile locations is often contemplated in papers that discuss the virtual data shadows to geographically bound situations such as the Arab Spring of 2011 or the Iran election protests of 2009 (Hecht et al. 2011; Lotan et al. 2011; Gaffney 2010). Takhteyev et al. (2011) attempted an automated coding of profile locations with an unnamed tool, but ultimately decided to hand code profile location details due to high-error rates. Vieweg et al. (2010) also handcoded profile locations, but also manually used tweet content in addition to profile content in determining the user's physical location. Java et al. (2007) used the Yahoo geocoding API, which attempts to assign a precise location to self-reported profile locations. However, the accuracy of such geocoding algorithms to profile location data on Twitter has not been previously determined. Most importantly, Hecht et al. (2011) show the need for great caution in finding that only 66 percent of the Twitter profiles they examined by hand had valid geographic information. 18 percent were blank and 16 percent had non-geographic information, mostly made of popular culture references. As a result, geocoding APIs will likely struggle with this input. In contrast to the free form nature of the profile location, Krishnamurthy et al. (2008) opt to use timezone (UTC offset) information in a user's profile to get a user's local time and thereby approximate longitude. Although it is impossible to determine latitude using this method, it remains that such a strategy can improve our best guesses about profile locations. However, it is unclear how many people actually set an accurate timezone. This is particularly a concern for users that employ third-party clients instead of visiting the Twitter website itself (within our sample fewer than 50 percent of tweets are created on Twitter's own website).

Newer research is developing methods to locate users based on the text content of their tweets, the time of day users tweet at, and/or the location of the users they are following or followed by (Cheng et al. 2010; Eisenstein et al. 2010; Hecht et al. 2011; Wing and Baldridge 2011; Mahmud 2012; Sadilek et al. 2012). All of these approaches, however, have only been developed and evaluated using tweets in the English language and/or geocoded tweets from the United States. This paper does not consider these developing approaches, but evaluates two off-the-shelf geocoding services and assesses their accuracy and performance across four different regions, only one of which is in the United States. In the manual examination of profile locations, the paper also raises hints of possible challenges these newer approaches will have to overcome to be geographically and linguistically broader. The paper also provides important insights into the disaccord between profile and device locations, which is important for the data used to develop and test these new approaches.

In sum, it is important to be aware of the myriad, overlapping and complex ways in which location is ascribed to information in Twitter before attempting to employ it in any geographic analyses. The following section more closely examines the accuracy and sources of error in a range of methods used to extract location and language from Twitter in order to build upon existing work and more clearly articulate how this information collected from Twitter might be of use for research in geography.

**Methods**
Between 10 November and 16 December, 2011, 111,143,814 tweets were collected by using the statuses/filter method of Twitter's streaming API[1] (these data are mapped in Figure 2). The method allows tweets to be collected from within a user-specified bounding box, which was drawn as a 180 by 360 degree box in this study (or a bounding box that encompasses the whole planet). Tweets sampled this way from the streaming API only include tweets with an explicit GeoIP or GPS 'device' location. The search API, which may make guesses as to users' location, was not used in this study. While some rate limiting errors are met when a large box is built (Twitter defines a maximum data rate, and any additional data above that limit is dropped from the stream), this effect is measurable. Rate limiting was only noticed during times that were co-incident with some North American weekday peak hours, and only 1.1 percent of our files were ultimately affected by any significant errors of data. Data were otherwise collected constantly with a few intermittent and brief crashes, and all downloaded information was stored in tab-separated files.

---

[1] This method captures tweets that are geocoded by both IP addresses and GPS-enabled devices.

**Figure 2**: Map of geotagged tweets captured between November 10 and December 16, 2011

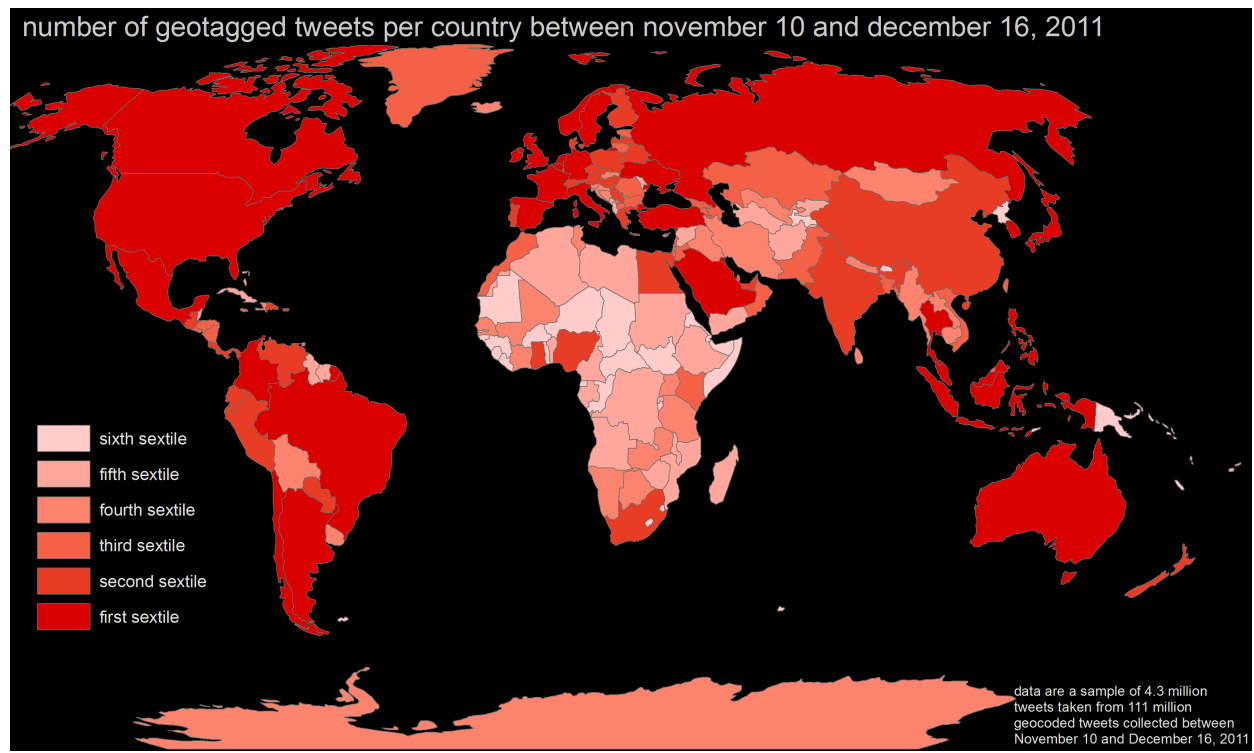

From our sample of 111 million tweets, 1,000 tweets were randomly selected from each of four metropolitan areas[2] (Cairo, Montreal, San Diego and Tokyo). These areas were selected by the research team as research sites that are characterized by interesting geographic, linguistic and cultural differences. To avoid over-representation by heavy users, we only include a maximum of one tweet per user in the sample. The location of every tweet was determined by the device location (GeoIP/GPS) recorded by Twitter.

Rather than relying on one particular algorithm, library or toolkit to establish a higher degree of accuracy, validity or certainty about the data, a central aim of this paper is to compare existing solutions. As such, the research team reviewed existing available geolocation and language identification solutions and evaluated them on their ease of use, throughput, and thoroughness. Based on these results, three language detection and two geolocation services were ultimately selected (Table 1).

Custom scripts employing each of the automated language identification algorithms (Alchemy, the Compact Language Detection kit and Xerox) and each of the geolocation solutions (Google and Yahoo) were written and the results of these algorithms were stored in a database. While other services certainly exist and the services considered in this paper are not exhaustive or authoritative, these services currently constitute the more easily implemented off-the-shelf solutions that are available. In particular, since Google switched its Language Detection API to a paid service, many researchers and companies have tried the CLD kit, but it has not been compared with other solutions. CLD is also forms part of the language-detection augmentation offered by DataSift, a Twitter data reseller, and hence is likely used by many commercial companies working with Twitter data. All of the language detections algorithms surveyed are pre-trained and immediately usable for any piece of text. This allows comparison apart from the specific data used to train the algorithm.

---

[2] The bounding boxes that we used to define the four urban areas are as follows: Cairo (31.1,29.95,31.54,30.28), Montreal (-74,45.33,-73.35,45.78), San Diego (-117.3,32.43,-116.74, 32.9), Tokyo (139.3,35.4,140.2,35.9). In all cases, outer ring roads were used to determine the approximate extent of each city's conurbation. While this approach is relatively imprecise, we deemed it equally problematic to establish a consistent bounding box size for all sample cities (due to the significantly different sizes of the four urban areas).

**Table 1**: Language and Location Services overview

| Service | Type | Use | Throughput | Thoroughness |
|---|---|---|---|---|
| Compact Language Detection Kit | Language Detection | C/C++ with a Ruby library and python bindings available | No limits: service is executed locally | 161 languages |
| Alchemy API | Language Detection | Web Service | 1,000 requests / day / API Key | 97 languages |
| Xerox Open Source | Language Detection | Web Service | No discernible limits | 46 languages |
| Twitter UI Language | User selected value | Delivered with tweet | n/a | 33 languages |
| Google Geocoding API | Geolocation | Web Service | 2,500 requests / day / IP | |
| Yahoo PlaceFinder | Geolocation | Web Service | 50,000 requests / day / API Key | |

In order to test language detection tools, we also randomly selected 1,000 tweets from each of our four study regions. These messages were then manually coded by the study's authors for the primary language of the tweet and disagreements were resolved through discussion. The study's authors collectively have experience with Arabic, English, German, Japanese, Korean, Mandarin, Persian, Spanish and Thai. Where words from multiple languages were found in a single tweet, the tweet was coded for the most abundant, primary language of the tweet.

**Findings**

*Language*

When examining the manual coding of language in 4,000 sample tweets, overall intercoder agreement was high between human coders (as shown in Table 2). Agreement was calculated with Fleiss' kappa. The measure ranges from -1, complete disagreement, to 1, perfect agreement, and, compared with percent agreement, better accounts for agreement between coders that could occur by simple chance. Conflicts between human coders were generally due to multiple languages being used within a single tweet. Tweets containing auto-generated text (e.g. automatically generated messages from Foursquare) often contained this mix of languages. In comparison to the human coders, the agreement between different language detection algorithms can be seen to be much lower (Table 2).

Table 2: Human and algorithm agreement on language (Fleiss' Kappa) with and without UI Language

| Research Site | Humans | | Machines | |
|---|---|---|---|---|
| | Coders | Fleiss' Kappa | Without UI Language | With UI Language |
| Cairo | 3 | 0.814 | 0.525 | - |
| Montreal | 3 | 0.779 | 0.587 | 0.461 |
| San Diego | 3 | 0.868 | 0.513 | 0.440 |
| Tokyo | 3 | 0.724 | 0.485 | 0.329 |
| Overall | 3 | 0.888 | 0.630 | 0.579† |

† Arabic was not available as a UI language choice at the time of the study; so, Cairo is not included in the overall statistics.

If the agreed human coding is treated as a 'gold standard,' the Compact Language Detection kit and Alchemy matched human classifications most closely, although all methods are with one standard deviation of each other (Table 3). Alchemy in general performed better than CLD with the significant exception of Tokyo. In the best case, Alchemy agreed with human coders on 91 percent of tweets in San Diego. However, even this only translates an intercoder agreement with a Fleiss' kappa of 0.644, which better accounts for agreement that could occur by simple chance as explained above. The best overall score was achieved by the CLD, which agreed with the human classification in 76.4 percent of cases. This translates to an intercoder agreement with a Fleiss' kappa of 0.670.

Yet, it remains that both of these scores are much lower than the overall intercoder agreement between the human coders, which had a Fleiss' kappa of 0.888. The relatively high percent agreement scores and lower kappa scores suggest disagreement between the algorithm and human coders occurred more about less frequently appearing languages. Nevertheless, the CLD is more apt for large datasets as it is the only local, fully offline method here considered. The code for CLD is also open-source and could be adopted, although the training corpora used to create the language identification fingerprints are unknown and unavailable.

Analysis shows the CLD performed particularly well in differentiating text in different Asian scripts. CLD and Alchemy did not do so well in Cairo where a number of Arabic-language tweets were written in the "Arabic chat alphabet" (i.e. Arabic using Latin characters). Whereas CLD nearly always classified text in Japanese, Korean, Chinese, and Arabic correctly when these languages were written in their usual scripts, it failed in all 89 cases to classify Arabic written with Latin characters correctly. Indeed, all of the language identification algorithms considered here failed to accurately classify these messages written in the Arabic chat alphabet.

It should also be pointed out that the user-interface language of Twitter users was a useful indicator of language in some research sites. It corresponded with the human coding of language for more than 75 percent of the tweets from Montreal, San Diego and Tokyo. At the time of data collection, there was not an option to set the Twitter user-interface language to Arabic, which likely explains why the user-interface language of users in Cairo only agrees with human coders for 45 percent of the tweets collected there.

Table 3: Algorithm agreement with human coders on language: Fleiss' Kappa/Scott's PI with percent agreement in parentheses. The standard deviation on percent agreement is about 0.5 in all cases.

| Research Site | Alchemy | Xerox | CLD | Twitter UI Lang |
|---|---|---|---|---|
| Cairo | 0.510 (69.6%) | 0.374 (55.9%) | 0.464 (61.7%) | -0.215 (44.9%)‡ |
| Montreal | 0.653 (83.1%) | 0.548 (73.9%) | 0.550 (74.7%) | 0.462 (75.6%) |
| San Diego | 0.644 (90.9%) | 0.501 (84.0%) | 0.487 (82.1%) | 0.565 (90.6%) |
| Tokyo | 0.017 (56.2%) | 0.029 (57.0%) | 0.418 (87.2%) | 0.278 (83.0%) |
| Overall | 0.609 (75.0%) | 0.534 (67.7%) | 0.670 (76.4%) | 0.714 (83.1%)† |

‡ Arabic was not available as a UI language choice at the time of the study; so, this value should be interpreted with caution.
† Arabic was not available as a UI language choice at the time of the study; so, Cairo is not included in the overall statistics.

Overall, language identification of tweets is difficult for human and machine coders alike. One preprocessing step that could improve results is to remove auto-generated text and non-language specific text (e.g. emoticons). It will be important to train machine algorithms on informal scripts (e.g. Arabic chat alphabet) in addition to classical scripts. The suitability of off-the-shelf language identification packages and the appropriateness of the user-interface language setting vary by research site; so, the best algorithm will likely depend on the specific research questions and study location.

*Geolocation*

To better understand how self-reported profile locations might be used to map the geography of information in Twitter, the Yahoo and Google geolocation algorithms were applied to 1,000 randomly selected users from each research site. Ideally, this study would apply the geolocation algorithms to the entirety of the users in each site, but rate limitations (i.e. the number of allowed requests per minute) prevent this from completing in a reasonable time frame. We found 16 percent of the location fields in our sample of 1,000 users from each research site were blank, which is similar to the 18 percent of profiles that Hecht et al. (2011) found blank in a general (geocoded and non-geocoded) sample of Twitter, although it is much lower than the 28 percent of profiles that Cheng et al. (2010) found blank in their general sample of Twitter profiles. Overall, Yahoo's and Google's geolocation algorithms perform similarly (Table 4). Excluding the blank profile locations, 94.5 percent of attempts using Yahoo's PlaceFinder and 86.2 percent of attempts using Google's Geocoder placed the user in some location. However many of these locations were outside of the bounding boxes defining each research site. On average, only 53.7 percent of attempts with Yahoo and 54.5 percent of attempts with Google placed the user within the bounding box from which they originally tweete. If blank profiles are included, these percentages drop to 45.0 percent for Yahoo and to 45.8 percent for Google, which would be closer to the likely upper bound on the actual percentage of users in a general sample from Twitter that could be placed correctly by geolocation algorithms alone.

Besides returning a geographic position, each geolocation service may report that it failed to geolocate the input. Comparing the values, it is clear that while Yahoo's algorithm geocodes more profile locations than Google's algorithm, many of these additional locations do not fall within the bounding boxes. Google's algorithm tends to fail to geocode a larger number of profile locations at all compared to Yahoo. However, of the locations Google's algorithm does geocode, more of these locations are within the bounding boxes of the research sites than Yahoo (Table 4). This is particularly apparent in the Tokyo research site where Google declines to determine a location for many more profile locations than Yahoo.

**Table 4:** Results of the geolocation of 1,000 randomly selected user profiles from each research site.

| Research Site | Blank | Yahoo | | | Google | | |
|---|---|---|---|---|---|---|---|
| | | In Box | Out of Box | Failed | In Box | Out of Box | Failed |
| Cairo | 201 | 431 | 336 | 32 | 444 | 295 | 60 |
| Montreal | 138 | 469 | 356 | 37 | 561 | 224 | 77 |
| San Diego | 133 | 446 | 359 | 62 | 407 | 348 | 112 |
| Tokyo | 170 | 456 | 335 | 39 | 419 | 197 | 214 |
| Overall | 642 | 1802 | 1386 | 170 | 1831 | 1064 | 463 |

Profile locations outside of the relevant bounding box may be due to users tweeting from a different location than that written in their profiles or due to geocoder error. To resolve this ambiguity, the authors manually examined all user locations that failed to geolocate or that geolocated outside of the relevant bounding box with both Google and Yahoo (Table 5). The largest portion (35.6 percent) of these profile locations were legitimate geographical locations outside of the bounding boxes. This suggests that users do not update their profile locations with great frequency. The second largest portion (24.1 percent) was non-geographic text (e.g. "Neverland") or generic, non-specific locations (e.g. Earth, a peach orchard). After this, a large portion (21.2 percent) was more general geographic locations that included the relevant research site (e.g. Japan, California or Kantou [the eastern half of Japan including Tokyo]).

The analysis also suggested ways to improve geolocation accuracy. Of the 5.8 percent of locations that were actually within the bounding boxes, abbreviations of place names was the most common reason for the geolocator to fail. Beyond this, another 4.2 percent of profile locations had multiple locations, one of which was the relevant study area. Finally, 9.1 percent of the tweets actually had latitude and longitude coordinates in the profile location field along with additional text (usually the name of an app placing the information in the profile location). Google and Yahoo recognized latitude and longitude coordinates without additional text, but any additional text causes both geocoders to fail (or in a smaller number of cases geolocate to a location that didn't correspond with the coordinates in the profile). All three of these situations, abbreviations, multiple locations, and latitude/longitude coordinates could likely be handled by preprocessing the data for these possibilities. This is especially applicable when targeting a single area, where a list of likely abbreviations might be more easily created.

**Table 5:** Human analysis of profiles failing to geolocate or geolocating outside of the relevant bounding box.

|  | Yahoo | Google |
|---|---|---|
| Profile location blank | 642 | 642 |
| Geolocated within bounding area | 1802 | 1831 |
| Geolocated outside bounding area | 1386 | 1064 |
|     Geolocated within bounding box by other algorithm | 206 | 107 |
|     Identified as within bounding area by human coder | 72 | 39 |
|     Not within bounding box (Human coder) | 481 | 462 |
|     More general (e.g. country, state, region) | 282 | 269 |
|     Multiple locations including within bounding area | 56 | 39 |
|     Latitude, Longitude pair | 21 | 10 |
|     Invalid / Generic (e.g. la la land, earth) | 268 | 138 |
| Failed to geolocate | 170 | 463 |
|     Geolocated within bounding box by other algorithm | 5 | 75 |
|     Identified as within bounding area by human coder | 4 | 37 |
|     Not within bounding box (Human coder) | 1 | 18 |
|     More general (e.g. country, state, region) | 4 | 16 |
|     Multiple locations including within bounding area | 1 | 18 |
|     Latitude, Longitude pair | 102 | 113 |
|     Invalid / Generic (e.g. la la land, earth) | 53 | 186 |

One vital caveat to these results is that the researchers coded data in the most naive form available. For both Yahoo's PlaceFinder and Google's Geocoder, additional options exist that may increase the accuracy of results. Yahoo's PlaceFinder returns multiple locations ordered by relevance for a given string (multiple locations can be, and are, returned routinely, such as Oxford, UK, and Oxford, Mississippi, when using a string of "Oxford"). This relevance, or confidence score, is between zero and one hundred and is shown with every returned result. This confidence score could be used to reject all results when every result is below a certain threshold. This would increase the number of locations that fail to geocode at all, but would likely raise the accuracy of profile locations that do geolocate. Furthermore, Google's API allows researchers to set a location hint in order to better capture data (potentially) originating from the region of focus.

Given the significant difficulties associated with geolocating profile locations, timezones have also been seen as a more reliable metric for approximating location (Krishnamurthy, Gill, and Arlitt 2008). Our dataset, however, suggests that many users have incorrectly configured timezones in their profiles (Table 6). Users select their timezone on the Twitter site from a predefined list. Several options have the same UTC offset (e.g. for UTC+9 a user can select "Tokyo" or "Seoul") although the day light saving or summer time rules may differ. It is likely that some users are traveling or have purposefully set different timezones from the locations in which they are using the service; however, the fact that only 57 percent of users tweeting in Montreal had set an east-coast timezone (much less the specific option of "Eastern Time (US & Canada)") indicates many users likely do not set their timezone correctly. In addition, many users tweeting from Montreal had other UTC+5 timezones selected (231 users in our sample of 1,000 users had set their timezone to Quito, for example), suggesting that caution is needed in interpreting the timezone more specifically than the UTC offset. Across all research sites 69.2 percent of users had selected a timezone with a UTC offset that corresponded to the device location information in the tweet (Table 6). The low number of users correctly setting their timezone may be influenced partially by the large number of 3rd party client devices used to tweet. Overall, only 23.6 percent of tweets captured across all our research sites were published via the Twitter website with the remainder sent using 3rd party applications. Geocoded tweets may be more likely to be sent from a 3rd party applications than non-geocoded tweets, however; so, the percentage of tweets sent from 3rd party applications in our sample of only geocoded tweets is likely higher than the percentage would be in a general sample of tweets including both geocoded and non-geocoded tweets.

Ultimately, our findings related to location point to the significant challenges associated with automatically identifying geographic references in unstructured text. It is important for researchers to be aware of these difficulties if they want to move beyond the limitations of only relying on the unrepresentative amount of information tagged with device locations.

**Table 6**: Timezone information has been seen as another proxy for location; however, this information is not routinely provided by all users.

| Research Site | All users captured | | | Geotagging user sample | | |
|---|---|---|---|---|---|---|
| | N | City-specific timezone | Correct UTC offset | N | City-specific timezone | Correct UTC offset midrule |
| Cairo | 1,952 | 54.6% | 55.4% | 1,000 | 55.5% | 56.3% |
| Montreal | 5,235 | 41.7% | 57.0% | 1,000 | 40.1% | 57.3% |
| San Diego | 9,292 | 57.1% | 60.5% | 1,000 | 58.6% | 63.2% |
| Tokyo | 55,573 | 68.2% | 72.3% | 1,000 | 70.1% | 73.8% |
| Overall | 72,052 | 64.4% | 69.2% | 4,000 | 56.1% | 62.7% |

**Discussion and Conclusions**

Over three hundred million users publish hundreds of millions of short messages every day on Twitter. As a result, this content has been used by researchers from fields as diverse as epidemiology, politics, marketing and geography to better understand, map and measure large-scale social, economic and political trends and patterns. However, much of this analysis is carried out with only limited understandings of how best to work with the spatial and linguistic contexts in which the information was produced. As such, it has been necessary to study the reliability of key methods used to determine language and location of content in Twitter.

This paper found that there are significant challenges to accurately determining the language of tweets in an automated manner[3]. None of the language identification methods tested in this paper is able to match the accuracy of human coding by multiple coders. The informal writing style, short length of tweets, use of multiple languages within a single tweet, and the presence of non-language specific content such as Uniform Resource Locators (URLs) and emoticons complicate the identification of language and limit accuracy.

The utility of the user-interface language setting varies across regions and languages. As of January 2013, Twitter has thirty-three user-interface languages available, which covers many major languages, but misses many key African, middle-eastern, Indian and Asian languages (e.g. Afrikaans, Zulu, Bengali, Marathi, Persian, Vietnamese and Javanese). The importance of these omissions will depend on the design of the study. It is important to note, however, that even when a user-interface language is present users writing primarily in that language may still not use that setting. This may be the case when a new language setting is introduced and user adoption lags or for font/device compatibility concerns (Warschauer, et al. 2002) among other reasons. The user-interface language setting also does not capture multilingual users who write in multiple languages on the platform (Eleta and Golbeck 2012).

Nevertheless, the CLD kit and Alchemy show useful promise as automated language identification packages. The former in particular has a great amount of flexibility as it can both be run offline and be modified as it is open-source. The best language identification algorithm for a particular study will depend on a number of factors. One important factor is the languages an algorithm is trained to identify and the scripts of these languages it is trained with (e.g. Arabic in Latin characters). CLD performed much better than Alchemy in Japan but otherwise Alchemy performed better in our other research sites. Yet, neither recognized the Arabic chat alphabet: a key omission that is likely to be mirrored in other informal and transliterated alphabets. Cases such as San Diego, however, show relative success in language detection. We may then conclude that on some level, the context of the study matters when considering algorithmic approaches to language identification.

Always running multiple language detection algorithms and reviewing subsets of the results with human coders, as in the work with Twitter of Carter et al. (2011) and the work with blogs of Hale (2012a), may give insight into the biases and reliability of different automated approaches and flag up potential issues on a specific sample of tweets. One method to potentially increase the accuracy of off-the-self language identification packages is to pre-process the tweets to temporarily remove emoticons, urls and other non-language specific text (something not attempted in this paper). Text automatically generated by third party services (e.g. Foursquare) often resulted in a mix of languages within a single tweet; so, identifying and temporarily removing this text could likely also increase the accuracy of off-the-self language identification packages. While removing such text temporarily may improve language identification, the text itself may nonetheless be useful for further analysis (e.g. studies of link diffusion across languages, e.g. Hale 2012b), and so researchers may wish to retain a copy of the unaltered tweet. Studying the effects of these pre-processing measures and the effects of using link content or grouping several tweets by a single user together for language identification as in Carter et al. (2011) will be a useful avenue for further research. The linguistic and geographic analysis of short, micro-blog texts is still an area of active research without any established best practices. Further studies to compare various methods and new approaches (such as crowdsourcing with Amazon's Mechanical Turk) are needed in order to identify concerns and possible future areas of improvement.

---

[3] It is our hope that future work will also consider the possibilities of using crowdsourced labor to accurately spatially reference tweets.

The paper also compared open-ended profile locations within four research sites in order to better understand how useful profile locations might be for studying the geography of information. Importantly, it finds that the geolocation results of profile locations are not a useful proxy for device locations (i.e. the place in which the information was disseminated), and identifies several reasons for this discord. This is an important finding not only for the social science analysis of where users are or perceive themselves to be, but also for computer science research which often uses geocoded tweets to evaluate the performance of new location classification approaches. For instance, Sadilek et al. (2012) demonstrate that when the location of a subset of users is known, it is then possible to infer the location of the friends of these users. The present work suggests that there will be an important difference in where users are placed depending on whether profile location or device location is used to create the starting set of users with a known location. The subset of users that geocode and the subset of users with clear place names in their profiles are unique, and importantly, this paper has found that even when both the profile and device location are valid, they do not always correspond. Similarly, there is a danger in relying on the device location as the baseline, true location of the user in training new geolocation algorithms based on text content (a practice used in, for example, Eisenstein et al. 2010; Wing and Baldridge 2011; Mahmud 2012).

The paper identifies three main reasons for the lack of correlation between profile and device location. First, commensurate with previous work (e.g. Hecht et al., 2011), a large number of profiles contained invalid, non-geographic text or simply larger geographic regions (countries, states). Secondly, adding to the literature, this paper finds, a large number of users tweeting within the study areas had profile locations set to locations outside of the study area – this likely resulted from users were commuting, traveling or simply not having updated their profile locations. Finally, several users were within the relevant study site but wrote their profile location information in such a way that the geolocation algorithms used failed to correctly code it. In addition to the recommendations by Hecht et al. (2011) to preprocess profile location information for fictitious names, this paper finds it is important to preprocess profile locations to handle abbreviations, lists of multiple locations, and latitude-longitude coordinates surrounded by other text. These steps should be investigated along with tweaking the available parameters to geolocation services. Several profiles had more general geographic boundaries (regions, states, countries), suggesting that the success at being able to place users within a geographic region will vary with the specificity of the region. Attempts to simply locate users within a country are more likely to be successful than trying to locate users to a specific city or metropolitan area. For city-level areas, local gazetteers might be useful (an approach not test here), but the analysis in this paper highlights the importance of supplementing such a list with common abbreviations, misspellings, other-language names, and transliterations of place names. Timezone information, specifically the Universal Time Coordinated offset (UTC-offset), while not perfect and showing differences in accuracy across study sites, seems to often correspond with the user's current location. UTC-offsets also have the value of being more easily processed than the free-form profile locations; however, UTC-offsets only give an indication of longitude and not latitude.

Because of the significant challenges associated with geolocating content and profiles in Twitter, it is tempting to associate certain languages with an assumed geographic origin of content. However, this paper demonstrates the large need for caution in using language as a proxy for location. Within each of the four research sites considered in this paper, a mix of languages was found suggesting that focusing on language as a proxy for location can lead to two issues. First such a strategy would miss other language users located within the location, and second, would likely capture users outside of the location of interest. Future work should look at the dispersion of various languages to determine to what extent language use clusters within certain geographic areas.

Although this paper highlights the challenges associated with accurately understanding the geography of information in Twitter, this should not lead us to discount the usefulness of profile locations as a means of geolocating content. Profile locations tell us much about how users perceive, present, and place themselves, and this paper has expanded on two methods that can be used to geolocate that unstructured information. Most importantly, the majority of the 300 million accounts on Twitter contain some type of profile location, whereas only a small proportion of tweets contain any structured device location. As such, further research and additional human coding of profile locations might be needed in order to accurately determine how well profile locations compare with device locations, how we might best geolocate profile locations, and the ways in which the geolocation of profile information might be linguistically or geographically contingent.

**Authors**

MARK GRAHAM is the Director of Research and a Research Fellow at the Oxford Internet Institute. He is also a Visiting Research Associate the University of Oxford's School of Geography and the Environment. His research focuses on Internet and information geographies, and the overlaps between ICTs and economic development.

SCOTT A. HALE is a research assistant and doctoral candidate at the Oxford Internet Institute, University of Oxford, interested in language separation online and the effects of platform design upon the transmission of information between speakers of different languages.

DEVIN GAFFNEY is senior developer at Little Bird in Portland, Oregon. He was previously affiliated with the Oxford Internet Institute as an MSc candidate and a Fell Fund grant-backed research assistant under Mark Graham. His research interests primarily focus on quantitative analyses and methodologies of social media data.

**References**

Bruns, A. and Burgress, J. E. 2011. #Ausvotes: How Twitter covered the 2010 Australian federal election. Communication, Politics and Culture 44: 37–56. http://eprints.qut.edu.au/47816/.

Carter, S., Tsagkias, M., and Weerkamp, W. 2011. Semi-supervised priors for microblog language identification. In Dutch-Belgian Information Retrieval Workshop (DIR 2011) http://wouter.weerkamp.com/downloads/dir2011-lid.pdf.

Cheng, Z., Caverlee, J., and Lee, K. 2010. You are where you tweet: A content-based approach to geo-locating Twitter users. In CIKM '10: 19th ACM International Conference on Information and Knowledge Management.

Eisenstein, J., O'Connor, B., Smith, N. A., and Xing, E. P. 2010. A latent variable model for geographic lexical variation. In EMNLP '10: 2010 Conference on Empirical Methods in Natural Language Processing, 1277–1287.

Eleta, I., and Golbeck, J. 2012. Bridging languages in social networks: How multilingual users of Twitter connect language communities. Proceedings of the American Society for Information Science and Technology, 49(1), 1-4. doi:10.1002/meet.14504901327.

Fischer, E. (2011). Language communities of Twitter. http://www.flickr.com/photos/walkingsf/6277163176/in/photostream.

Fleiss, J. L. 1971. Measuring nominal scale agreement among many raters. Psychological Bulletin, 76(5): 378-382.

Gaffney, D. 2010. #iranElection: Quantifying online activism. In Proceedings of Web Science 10: Extending the Frontiers of Society On-Line. Raleigh, NC, USA: Web Science Trust. http://journal.webscience.org/295/.

Graham, M. and Zook, M. 2011. Visualizing global cyberscapes: Mapping user-generated placemarks. Journal of Urban Technology 18: 115–132.

Gruzd, A., Wellman, B., and Takhteyev, Y. 2011. Imagining Twitter as an imagined community. American Behavioral Scientist 55: 1294–1318.

Gwet, K. L. 2010. Handbook of inter-rater reliability. 2nd edition. Gaithersburg: Advanced Analytics.

Hale, S. A. 2012a. Net Increase? Cross-lingual linking in the blogosphere. Journal of Computer-Mediated Communication, 17(2): 135-151.

——— 2012b. Impact of platform design on cross-language information exchange. In Proceedings of the 30th International Conference on Human Factors in Computing Systems, CHI '12, ACM

Halley, E. 1731. A proposal of a method for finding the longitude at sea within a degree, or twenty leagues. Philosophical Transactions (1683-1775) 37 (January 1, 1731): 185–195.

Hecht, B., Hong, L., Suh, B., and Chi, E. 2011. Tweets from Justin Bieber's heart: The dynamics of the location field in user profiles. In Proceedings of the 2011 Annual Conference on Human Factors in Computing Systems, 237–246. New York, NY, USA: ACM.

Honeycutt, C. and Herring, S.C. 2009. Beyond Microblogging: Conversation and collaboration via Twitter. In System Sciences, 2009. HICSS '09. 42nd Hawaii International Conference on System Sciences (HICSS-42), 1–10.

Hong, L., Convertino, G., and Chi, E. 2011. Language matters in Twitter: A large scale study. In International AAAI Conference on Weblogs and Social Media, 518–521.

Krishnamurthy, B., Gill, P., and Arlitt, M. 2008. A few chirps about Twitter. In Proceedings of the First Workshop on Online Social Networks, 19–24. New York, NY, USA: ACM.

Lotan, G., Graeff, E., Ananny, M., Gaffney, D., Pearce, I., and Boyd, D. 2011. The revolutions were tweeted: Information flows during the 2011 Tunisian and Egyptian revolutions. International Journal of Communication, 5: 1375-1405.

Mahmud, J. 2012. Where is this tweet from? Inferring home locations of Twitter users. In ICWSM '12: Sixth International AAAI Conference on Weblogs and Social Media.

Palen, L., Vieweg, S., and Anderson, K. M. 2011. Supporting 'everyday analysts' in time- and safety- critical situations. The Information Society Journal, 27(1): 52-62.

Romero, D. M., Meeder, B., and Kleinberg, J. 2011. Differences in the mechanics of information diffusion across topics: Idioms, political hashtags, and complex contagion on Twitter. In Proceedings of the 20th International Conference on World Wide Web, 695–704. New York, NY, USA: ACM.

Sadilek, A., Kautz, H., and Bigham, J. P. 2012. Finding your friends and following them to where you are. In WSDM '12: Fifth ACM International Conference on Web Search and Data Mining, WSDM '12., 723–732.

Semiocast. 2010. Half of messages on Twitter are not in English: Japanese is the second most used language. Semiocast Press Release, Paris, France.

Shelton, T., Zook, M., and Graham, M. 2013. The technology of religion: Mapping religious cyberscapes. The Professional Geographer, 65. doi:10.1080/00330124.2011.614571.

Takhteyev, Y., Gruzd, A., and Wellman, B. 2011. Geography of Twitter networks. Social Networks: 1–26. doi:10.1016/j.socnet.2011.05.006.

Vieweg, S., Hughes, A. L., Starbird, K., and Palen, L. 2010. Microblogging during two natural hazards events: What Twitter may contribute to situational awareness. Proceedings of the 28th International Conference on Human Factors in Computing Systems (pp. 1079-1088). New York, NY, USA: ACM.

Warschauer, M., Said, G. R. E., & Zohry, A. 2002. Language choice online: Globalization and identity in Egypt. Journal of Computer-Mediated Communication, 7(4). Retrieved from http://jcmc.indiana.edu/vol7/issue4/warschauer.html.

Wing, B. P., and Baldridge, J. 2011. Simple supervised document geolocation with geodesic grids. In ACL '11: Proceedings of the 49th Annual Meeting of the Association for Computational Linguistics, 955–964. Portland, OR.

Zook, M., Graham, M., Shelton, T., and Gorman, S. 2010. Volunteered geographic information and crowdsourcing disaster relief: A case study of the Haitian earthquake. World Medical & Health Policy 2 (Jul): 7–33.